\documentclass[prb,superscriptaddress,twocolumn,showpacs]{revtex4}
\usepackage[english]{babel}
\usepackage{amsmath,amssymb,amsfonts}
\usepackage{graphicx}

\begin{document}
\title{Cotunneling in the \boldmath{$\nu=5/2$} fractional quantum Hall regime}

\author{Robert Zielke}
  \affiliation{Department of Physics, University of Basel,
    Klingelbergstrasse 82, 4056 Basel, Switzerland}
\author{Bernd Braunecker}
  \affiliation{Departamento de F\'isica Te\'orica de la Materia
    Condensada, Facultad de Ciencias, Universidad Aut\'onoma de
    Madrid, 28049 Madrid, Spain}
\author{Daniel Loss}
  \affiliation{Department of Physics, University of Basel,
    Klingelbergstrasse 82, 4056 Basel, Switzerland}
\date{\today}

\begin{abstract}
We show that cotunneling in the 5/2 fractional quantum Hall regime
allows us to test the Moore-Read wave function, proposed for this regime,
and  to probe the nature of the fractional charge carriers.  We
calculate the cotunneling current  for electrons that tunnel between
two quantum Hall edge states via a quantum dot and for quasiparticles
with fractional charges $e/4$ and $e/2$ that tunnel via an antidot.
While electron cotunneling is strongly suppressed, the quasiparticle
tunneling shows signatures characteristic of the Moore-Read state.
For comparison, we also consider  cotunneling between Laughlin states,
and  find that  electron transport between Moore-Read states and
between Laughlin states at filling factor 1/3 have identical voltage
dependences.
\end{abstract}


\maketitle

\section{Introduction}
Fractional quantum Hall (FQH) states are intriguing states of matter
because elementary collective excitations behave as quasiparticles
with fractional charge and statistics.  The FQH state at filling
factor 5/2 (5/2-FQHS) has become of special interest because it has
been identified in several proposals as a state in which the
elementary excitations obey non-Abelian fractional statistics.\cite{
  moore_nonabelions_1991,fradkin_chern-simons_1998,read_beyond_1999,
  read_paired_2000,lee_particle-hole_2007,levin_particle-hole_2007,
  nayak_non-abelian_2008}  Numerical simulations testing these
proposals have remained inconclusive,
\cite{morf_transition_1998,rezayi_incompressible_2000,feiguin_density_2008,
feiguin_spin_2009,moller_paired_2008,zozulya_entanglement_2009}
mainly due to finite size limitations. A proof of the nature of the
5/2-FQHS should therefore come from experiments.\cite{
  radu_quasi-particle_2008}  As a first indicator, evidence for a
chiral Luttinger liquid at the edges of the FQH sample was obtained in
Ref. \onlinecite{miller_fractional_2007}, demonstrating the fractional
nature of the quantum state, yet not its Abelian/non-Abelian
statistics.  It has also been shown that quantum point contacts and
interferometers can be constructed in the samples,\cite{
  ji_electronic_2003,schoenenberger_interference_1999} allowing for
the implementation of the interferometer-based tests proposed in
Refs. \onlinecite{law_electronic_2006,feldman_detecting_2006,feldman_shot_2007,
  wang_identification_2010}. A thermoelectric probing of different FQH
states on  quantum dots has been proposed in
Ref. \onlinecite{viola_thermoelectric_2012}.  Yet further proposals
for tests are desirable to obtain conclusive evidence.  Theoretical
FQH studies have shown that quantum dot (QD) and  quantum antidot (AD)
structures with corresponding excitations,  electrons, and
quasiparticles (QPs), exhibit similar physics.\cite{
  de_c._chamon_two_1997,geller_aharonov-bohm_1997}  In the cotunneling
regime the number of particles on the dot is  conserved and second
order tunneling processes dominate transport.\cite{
  averin_single_1992,golovach_transport_2004} Elastic (inelastic)
processes conserve (change) the state of the dot.  The inelastic
process leads to an excitation of the dot for bias voltages larger
than the level spacing on the dot.  Since the implementation of dot
structures has become a mature  experimental technique, we propose in
the present  paper a QD based setup for an alternative test of the
nature of the  5/2-FQHS. In particular, we show that  the cotunneling
current strongly  depends on the nature of the elementary QP
excitations that  can contribute to the current that is allowed to
tunnel through the dots.

Possible charge carriers are electrons and fractionally charged QPs
with non-Abelian statistics.  The most prominent candidate QPs are
excitations of charge $e/4$ and $e/2$ with $e$ being the electron
charge.\cite{
  fendley_edge_2007,bishara_edge_2008,dolev_observation_2008,
  venkatachalam_local_2011,carrega_anomalous_2011,carrega_spectral_2012}
We investigate the existence of signatures of the 5/2-FQHS according
to the theoretical description proposed by Moore and Read (MR)
\cite{moore_nonabelions_1991} in simple transport measurements,
e.g., conductance through a QD.

\begin{figure}[t]
\includegraphics[width=.475\textwidth]{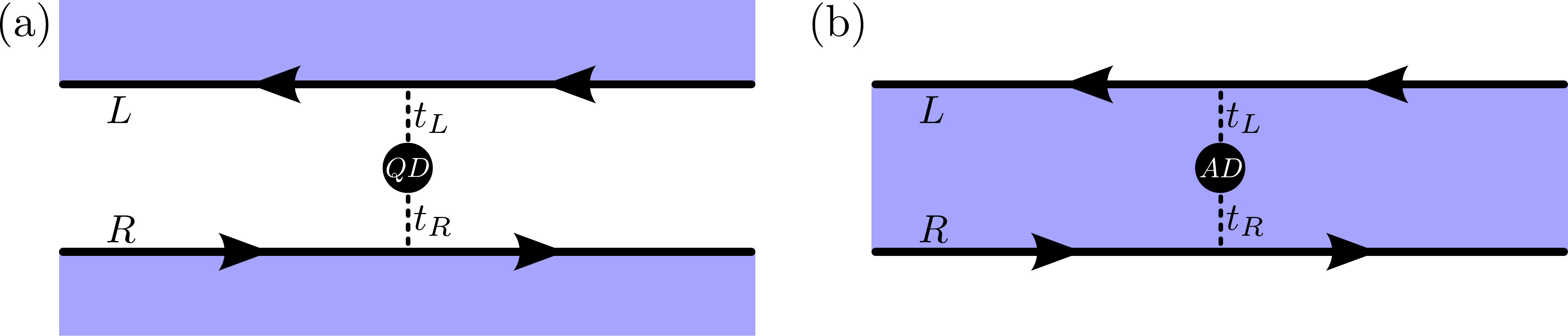}
\caption{Sketch of the two setups considered in this work: Moore-Read
  edge states $L$ and $R$, at chemical potentials $\mu_{L,R}$, tunnel
  coupled to (a) a quantum dot (QD) and (b) a quantum antidot (AD).
  The shaded (blue) region represents the bulk of the FQH samples.  In
  setup (a) tunneling is limited to electrons, while in  (b) the
  quasiparticles (QP) that can tunnel are determined by the filling
  factor of the AD, and, for the MR state both, charge-$e/2$ and
  -$e/4$, QPs are possible.}
\label{fig:setup}
\end{figure}
Figure 1 shows the two situations of interest: (a) two different FQH
samples at filling factor $\nu=5/2$ which are weakly tunnel coupled to
a QD, and (b) one single FQH sample whose edge states are weakly
tunnel coupled to an AD in the bulk.  The tunneling particles in the
latter are non-Abelian QPs instead of electrons.  The edge states are
modeled by a chiral Luttinger liquid theory
\cite{wen_chiral_1990,wen_theory_1992,wen_topological_1995,
  giamarchi_quantum_2004,gogolin_bosonization_2004} corresponding to
the MR state whose eigenmodes are fractional excitations.  In this
work we focus on cotunneling in the Coulomb blockade regime close
above a sequential tunneling peak such that it is energetically
favorable to first remove a particle from the dot rather than first
adding another particle.  An interesting outcome of our calculation is
that electron cotunneling via a QD between both Laughlin edge states
at filling factor $\nu=1/3$ and MR edge states shows the same bias
voltage dependence.

\section{Result}
The cotunneling current from lead $l$ to lead $l'$ in lowest order in
the bias $V$ is given by $I=I_{el}+I_{inel}$, with $I_{el}$ being the
elastic cotunneling current given by
\begin{equation}
\label{eq:I_el}
I_{el}=
\frac{2\pi}\hbar
\frac{\left(\frac{V}\Lambda\right)^{2\kappa-2}}{\Gamma(2\kappa)}
\frac{\gamma_{l}\gamma_{l'}V}{(\mu_l-\varepsilon_{nq})^2}
\ \theta(V),
\end{equation}
and $I_{inel}$ the inelastic cotunneling current given by
\begin{equation}
\begin{split}
\label{eq:I_inel}
I_{inel}=&\ 
\frac{2\pi}\hbar
\frac{\left(\frac{V-\Delta}\Lambda\right)^{2\kappa-2}}{\Gamma(2\kappa)}
\frac{\gamma_{l}\gamma_{l'}(V-\Delta)}{(\mu_l-\varepsilon_{nq})^2}
(1+\rho^*)\ \theta(V-\Delta)
,
\end{split}
\end{equation}
with bias $V=\mu_{l}-\mu_{l'}$, chemical potential $\mu_l$ of edge
state $l$, single-particle level spacing on the dot
$\Delta=\varepsilon_{n'q}-\varepsilon_{nq}$, dot level
$\varepsilon_{nq}$, effective bandwidth of the leads $\Lambda=\hbar
u/\alpha$ (bounded by the gap of the 5/2-FQHS) defining  the length
$\alpha$, velocity of the bosonic (fermionic) edge excitations
$u\ (v_n)$, Heaviside step function $\theta$, and tunneling rate
$\gamma_{l}=|A_{ln'n}|^2/(2\pi v_n\hbar)$ with $A_{ln'n}=\sum_p
t_{lp}\langle n'|d_{pq}|n\rangle$ (see below).  The renormalized dot
occupation
$\rho^*\propto(V+\Delta)^{2\kappa-1}-(V-\Delta)^{2\kappa-1}$ accounts
for the overshooting of the conductance close to the transition from
the elastic to the inelastic regime and will be given in detail below.
The parameter $\kappa$ is determined by the type of tunneling
particles (see below).  For cotunneling of $e/2$ and $e/4$ QPs in
setup (b), we replace the step function $\theta$ by the Fermi function
$f(V)=\left(1+e^{V/k_BT}\right)^{-1}$ (for a temperature
$T<\Delta/k_B$ and the Boltzmann constant $k_B$) which smooths out
the discontinuities at $V=0,\Delta$ for $T=0$, and we use the
$V\rightarrow\sqrt{V^2+(k_BT)^2}$  regularization for cotunneling of
$e/4$ QPs.

\section{Model}
The system is modeled by the Hamiltonian $H=H_0+H_T$, where
$H_0=H_L+H_R+H_D$ describes the uncoupled FQH edges and the dot, and
$H_T$ the tunneling between them.  In the considered systems the leads
are fractional quantum Hall edge states modeled according to the MR
state and described by the Hamiltonian for lead $l$ at chemical
potential $\mu_l$ \cite{moore_nonabelions_1991,fendley_edge_2007}
\begin{equation}
\label{eq:lead_hamiltonian}
H_{l}=\frac{u \hbar}{2\pi}\int dx\big(\partial_x\phi_l(x)\big)^2
-i v_n\hbar\int dx\;\eta_l(x)\partial_x\eta_l(x),
\end{equation}
where $\phi_l$ is a chiral boson field, the Majorana field $\eta_l$ is
the zero mode, and $u$ ($v_n$) is the velocity of the bosonic (neutral
fermionic) excitations. For the MR state the lower filled Landau level
acts as a background potential and causes a shift of the energy levels
which is not important in our discussion.  The fermion operator in
lead $l$ is given by \cite{fendley_edge_2007,wen_topological_1995}
\begin{equation}
\label{eq:fermion-operator}
\psi_{le}(x,t)=\frac{e^{-ik_1x}}{\sqrt{2\pi \alpha}}
\ \eta_l(x,t)
\ e^{i\phi_l(x,t)\sqrt{2}},
\end{equation}
with $k_1$ proportional to the particle density in the leads.
Analogously the $e/2$ and $e/4$ QP operators are
\cite{fendley_edge_2007,wen_topological_1995,fendley_dynamical_2006}
\begin{align}
\label{eq:half-charge-operator}
\psi_{l\frac e2}(x,t)=&\ \frac{e^{-ik_1x}}{\sqrt{2\pi \alpha}}
\ e^{i\phi_l(x,t)/\sqrt{2}},
\\
\label{eq:quarter-charge-operator}
\psi_{l\frac e4}(x,t)=&\ \frac{e^{-ik_1x}}{\sqrt{2\pi \alpha}}
\ \sigma_l(x,t)\ e^{i\phi_l(x,t)/2\sqrt{2}},
\end{align}
where $\sigma_l$ is a chiral Ising spin field.  The dot is modeled by
$H_{D}=\sum_{n}\varepsilon_{nq} d^\dagger_{nq} d_{nq}$, where $d_{nq}$
is the electron or QP operator for the discrete particle level $n$ on
the dot similar to Eqs. \eqref{eq:fermion-operator},
\eqref{eq:half-charge-operator}, and
\eqref{eq:quarter-charge-operator} with particle charge $q=e$ for
electrons on the QD and $q=e/2,e/4$ for QPs on the AD.  The Coulomb
repulsion of the particles lifts a possible degeneracy of the energy
levels including both charging and interaction energies.  Tunneling
between the leads and the dot is described by the perturbation
$H_T=\int dx\sum_{l,n}t_{ln}\psi_{lq}^\dagger(x) d_{nq}+h.c.$, where
$\psi_{lq}(x)$ and $d_{nq}$ annihilate particles of charge $q$ in lead
$l$ at $x$ and in the dot level $n$, respectively, and with tunneling
amplitude $t_{ln}$.  From Ref.  \onlinecite{fendley_edge_2007} we know
that QP tunneling processes are dominant in the MR state.

In setup (a) (Fig. \ref{fig:setup}) the leads are independent as they
belong to different FQH systems.  Independence between the edges is
also assumed in setup (b), which requires that the length of the edges
be larger than the coherence length of the excitations, estimated as
$2.3 l_0$ with $l_0\sim 4\mu m$ being the magnetic length.\cite{
  baraban_numerical_2009,wan_fractional_2008}  Tunneling in the leads
is assumed to be limited to the positions closest to the dot, denoted
by $x=0$ with width $\Delta x\ll {k_1}^{-1}$, because $t_{ln}$ depends
exponentially on the tunneling distance
\cite{recher_superconductor_2002} and consequently no special effects
arise from the difference in velocities between bosons and Majorana
states as opposed to Mach-Zehnder interferometers.\cite{
  law_electronic_2006,feldman_detecting_2006,feldman_shot_2007,
  wang_identification_2010}  We focus on the Coulomb blockade regime,
where the particle number on the dot is fixed, $N_q=\sum_n
d^\dagger_{nq}d_{nq}$.  The charging energy of the dot is much larger
than the single-particle level spacing $\Delta$ and particularly
larger than $\mu_l-\varepsilon_{nq}$.  Tunneling processes through
energetically distant dot levels are suppressed.  The dot forms an
effective two-level system that contains one particle.  The
persistence of the ground state in the elastic regime is ensured by
charge conserving cotunneling processes which relax the state of the
dot.  The charge of the tunneling QPs is set by the filling factor of
the AD such that there is no mixing of $e/4$ and $e/2$ QPs.  We
consider a regime in which the applied bias and the level spacing on
the dot are larger than the temperature, allowing us to essentially
neglect temperature effects.  Since the MR state is spin-polarized,
spin is neglected in the model
\cite{morf_transition_1998,rezayi_incompressible_2000,feiguin_spin_2009,
  chesi_quantum_2008,tiemann_unraveling_2012,stern_nmr_2012} and the
fermion operator $\psi_{le}$ is identified with the spin-polarized
electron.

\section{Transition rates}
The transition rate $W_{l'l}(n',n)$ of 
transferring a particle from lead $l$ to lead $l'$ and shifting the
particle on the dot from level $n$ to level $n'$ is determined by the
golden rule
\begin{equation}
\label{eq:golden-rule-2}
W_{l'l}(n',n)
=
\frac{2\pi}\hbar
\sum_{|F\rangle\ne |I\rangle}
\big|\langle F|\hat T|I\rangle\big|^2
\delta(E_I-E_F),
\end{equation}
with the T matrix $\hat T=H_T (E_I-H_0+i0^+)^{-1}\hat T$, final,
$|F\rangle$, and initial, $|I\rangle$, states for the two leads and
the dot, and energies $E_F$, $E_I$ respectively.  Level broadening in
the leads (reservoirs) is neglected in state $|F\rangle$.  We focus on
the second order in the tunneling Hamiltonian $H_T$ and the
cotunneling regime.  The linear order of $H_T$ in the T matrix leading
to sequential tunneling is suppressed in the Coulomb blockade regime
\cite{averin_single_1992} such that only the next order, cotunneling,
matters.  The final and initial states are given by
$|F\rangle=|F_{l}\rangle\otimes|F_{l'}\rangle\otimes|F_D\rangle$ and
$|I\rangle=|I_{l}\rangle\otimes|I_{l'}\rangle\otimes|I_D\rangle$, with
$|F_{l}\rangle=\psi_{lq}\ |I_{l}\rangle$,
$|F_{l'}\rangle=\psi_{l'q}^\dagger\ |I_{l'}\rangle$,
$|F_{D}\rangle=d_{n'q}^\dagger d_{nq} |I_{D}\rangle$, and initial lead
$l$ and dot states $|I_{l}\rangle$ and $|I_{D}\rangle$, respectively.

From Eqs.~\eqref{eq:fermion-operator},~\eqref{eq:half-charge-operator},
and \eqref{eq:quarter-charge-operator} we obtain the correlation
functions for electrons, $q=e/2$, and $q=e/4$ QPs, respectively,
\cite{wen_topological_1995}
\begin{align}
\label{eq:corr-e}
\langle\psi_{le}^\dagger(t)\psi_{le}(0)\rangle
=&\
\frac{\langle\eta_l(t)\eta_l(0)\rangle}{2\pi \alpha}
\ e^{2 J(t)}
\propto t^{-(\kappa=3)},
\\
\label{eq:corr-e/2}
\langle\psi_{l\frac e2}^\dagger(t)\psi_{l\frac e2}(0)\rangle
=&\
\frac1{2\pi \alpha}
\ e^{J(t)/2}
\propto t^{-(\kappa=1/2)},
\\
\label{eq:corr-e/4}
\langle\psi_{l\frac e4}^\dagger(t)\psi_{l\frac e4}(0)\rangle
=&\
\frac{\langle\sigma_l(t)\sigma_l(0)\rangle}{2\pi \alpha}
\ e^{J(t)/8}
\propto t^{-(\kappa=1/4)},
\end{align}
with $J(t)=-\ln\big((-u t+i\alpha)/i\alpha\big)$.  The indicated
power laws define the coefficient $\kappa$.  The noninteracting case,
$\kappa=1$, corresponds to the Fermi liquid (FL) limit.  We note that,
since $x=0$, the Majorana fermion field $\eta_l$ (Ising spin field
$\sigma_l$) results in a simple correlator, increasing the lead
correlation exponent $\kappa$ by $1$ ($1/8$) such that
$\kappa=3,1/2,1/4$ for $q=e,e/2,e/4$, which is in contrast to the
Mach-Zehnder interference proposals
\cite{law_electronic_2006,feldman_detecting_2006,feldman_shot_2007,
  wang_identification_2010} and leads to simpler propagators.

In the inelastic cotunneling regime, $V>\Delta$, processes exist which
excite the dot and change the occupation probabilities of the dot
levels.  The steady state occupation probabilities are then determined
by the master equation $W^\uparrow \rho(1)-W^\downarrow \rho(2)=0$,
with $W^\uparrow=\sum_{l,l'}W_{l'l}(2,1)$, 
$W^\downarrow=\sum_{l,l'}W_{l'l}(1,2)$, and $\rho(n)$ the occupation
probability of level $n=1,2$.  The cotunneling current from lead $l$
to lead $l'$ is then given by
\begin{equation}
\label{eq:master-equation}
I_{l'l}=\sum_{n,n'}W_{l'l}(n',n) \rho(n).
\end{equation}
The renormalized dot occupation in Eq. \eqref{eq:I_inel} reads
$\rho^*=(W_{l'l}(1,2)-W_{l'l}(2,1))/(W^\uparrow+W^\downarrow)$.
\begin{figure}[t]
\includegraphics[width=.475\textwidth]{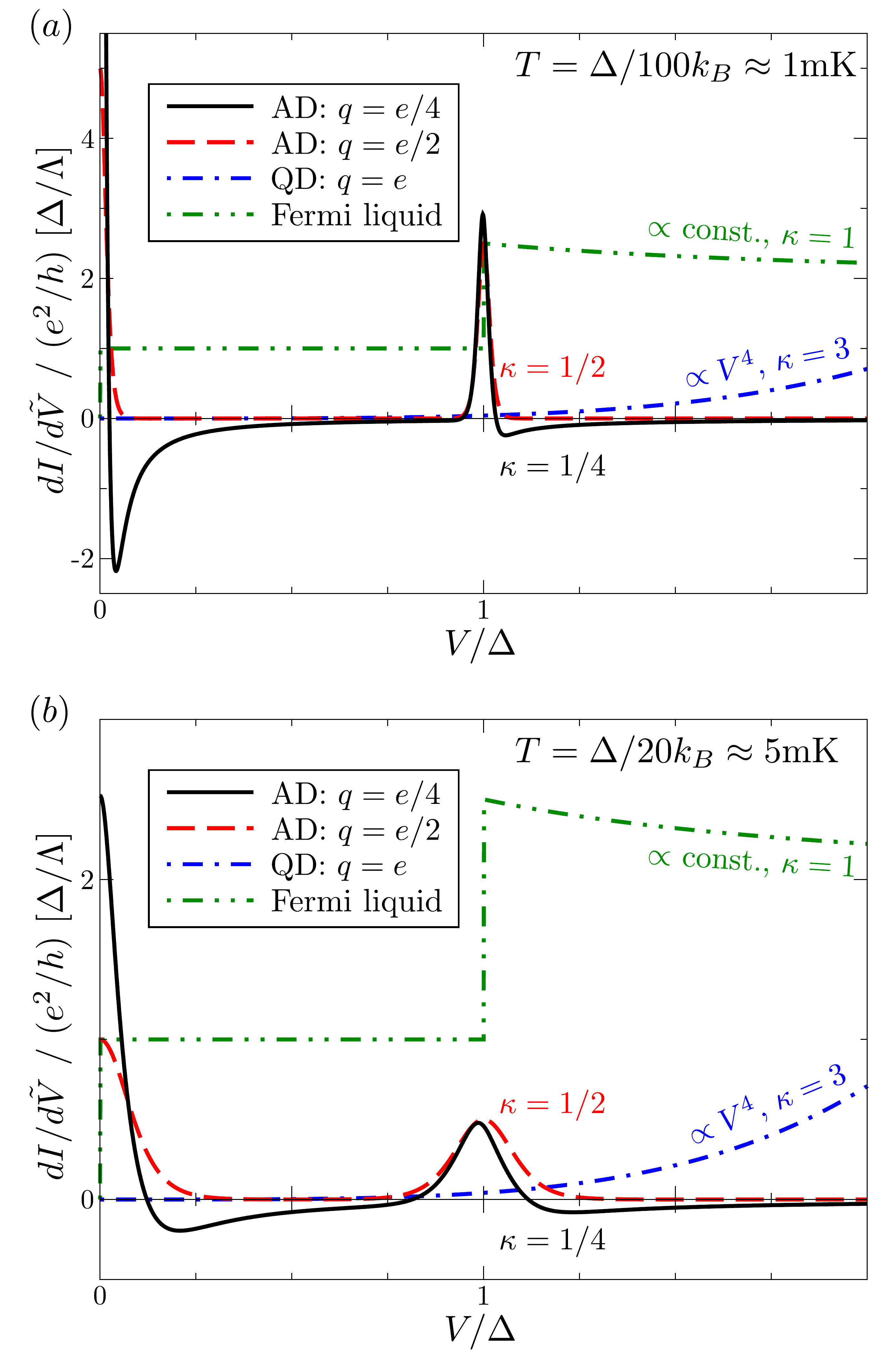}
\caption{Cotunneling conductance $dI/d\tilde V$ with $\tilde V=V/e$,
  and $I=I_{el}+I_{inel}$ given by Eqs. \eqref{eq:I_el} and
  \eqref{eq:I_inel} for $e/4$ and $e/2$ QP transport between MR edge
  states of the same FQH sample through an AD, electron transport
  between MR edge states of separate FQH samples through a QD, and
  electron transport between FL leads through a QD.  The $e/2$ and
  $e/4$ QP conductance is shown for the temperatures (a)
  $T=\Delta/100 k_B\approx1\mbox{mK}$ and (b)  $T=\Delta/20
  k_B\approx5\mbox{mK}$, respectively.}
\label{fig:fermi-e/4-e-cond}
\end{figure}
We evaluate Eq. \eqref{eq:golden-rule-2} by means of Fourier
integration over time  to reexpress the T matrix in terms of the
correlators given in Eqs. \eqref{eq:corr-e}-\eqref{eq:corr-e/4}. We
then find for the cotunneling rates for charge transfer between FQH
edge states and a dot
\begin{equation}
\label{eq:Wll'}
W_{l'l}(n',n)=
\frac{2\pi}\hbar
\ \Gamma_{l'ln'n}^{\kappa}\
\frac{\gamma_{l}\gamma_{l'}(V-\Delta)}{(\mu_l-\varepsilon_{nq})^2}
\ \theta(V-\Delta),
\end{equation}
with the real coefficients
\begin{equation}
\label{eq:Gamma}
\Gamma_{l'ln'n}^{\kappa}=
\left.\frac{\left(\frac{V-\Delta}\Lambda\right)^{2\kappa-2}}{\Gamma(2\kappa)}
\right|_{l\ne l'}
+\delta_{l',l}
\ \Xi^{\kappa}_{ln'n},
\end{equation}
where $\Gamma$ is the Gamma function and $\Xi^{\kappa}_{ln'n}$ results
from energy renormalization in the leads.  In the limit of
noninteracting leads, $\Gamma_{l'ln'n}^{1}=1$ and hence
$\Xi^{1}_{ln'n}=1$, the FL result of
Ref. \onlinecite{recher_quantum_2000}  close to a sequential tunneling
resonance  is recovered.  For charge conserving processes of $e/2$
($e/4$) QPs, $\Xi^{1/2}_{ln'n}$ ($\Xi^{1/4}_{ln'n}$) can be
approximated by their most singular contributions, the branch points
(branch cuts), such that $\Xi^{1/2}_{ln'n}=-\Lambda/\Delta$
($\Xi^{1/4}_{ln'n}=|\Lambda/\Delta|^{3/2}/\sqrt\pi$), whereas for
electron cotunneling processes $\Xi^{3}_{ln'n}$ results in a lengthy
expression with, however, negligible effect. From
Eqs. \eqref{eq:master-equation}-\eqref{eq:Gamma} we obtain the results
of Eq. \eqref{eq:I_el} (for $\Delta=0$) and Eq. \eqref{eq:I_inel}.

\begin{figure}[t]
\includegraphics[width=.475\textwidth]{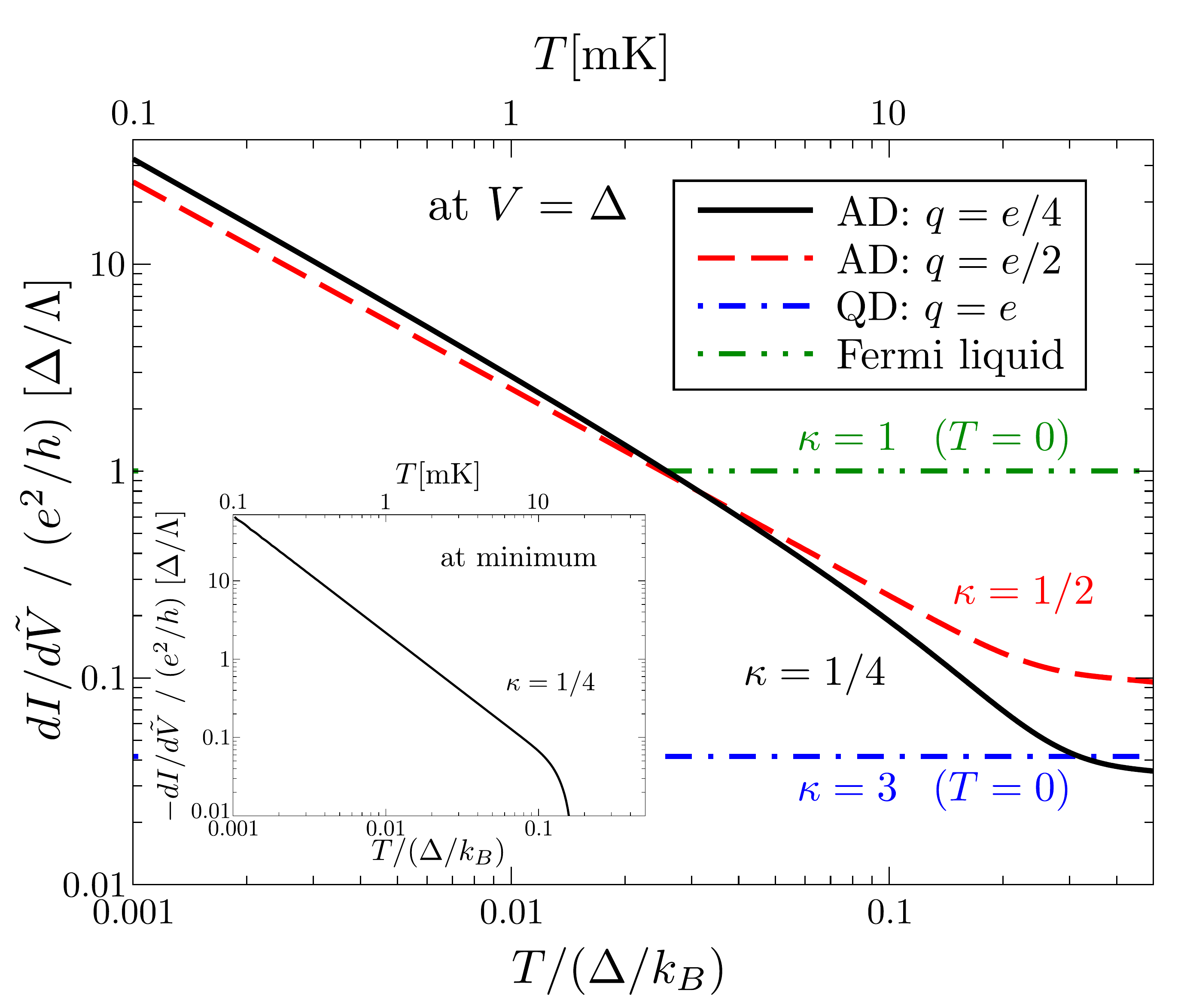}
\caption{Temperature dependence of the $e/4$ and $e/2$ QP cotunneling
  conductance $dI/d\tilde V$ with 
  $I=I_{el}+I_{inel}$  given by Eqs. \eqref{eq:I_el} and
  \eqref{eq:I_inel} and $\tilde V=V/e$ fixed at the
  transition from the elastic to the inelastic regime, $V=\Delta$.
  Values for electron transport between
  MR edge states of separate FQH samples through a QD and electron
  transport between FL leads through a QD are shown for zero
  temperature. On the upper scales $T$ is given in mK for
  $\Delta/k_B=100\mbox{mK}$. The inset shows the conductance dip 
  depth, $-dI/d\tilde V$, for $e/4$ QP tunneling at the minimum 
  in the nonfixed bias $V$.  }
\label{fig:peak-temperature-dependence}
\end{figure}

\section{Discussion}
We have calculated the cotunneling current of the MR state in both
setups of electron and quasiparticle tunneling via a dot.  The
resulting line shapes differ significantly, especially from the line
shape of the FL regime.\cite{golovach_transport_2004}  The cotunneling
current for electrons shows a power law dependence on both the bias
applied to the edge states and the effective bandwith of the leads,
$V^{2\kappa-1}$ and $\Lambda^{2-2\kappa}$.  The effective bandwidth
$\Lambda$ is on the order of the effective Landau level gap size in
the 5/2-FQHS, $\sim$ 100-500 mK. We note that
$\bar\gamma_l=\gamma_l\ \Lambda^{1-\kappa}$ can be considered an
effective cotunneling amplitude from which it is obvious that electron
cotunneling between MR edge states via a QD is highly suppressed by
the fourth inverse power of the effective bandwidth,
$\bar\gamma_L\bar\gamma_R=\gamma_L\gamma_R\ \Lambda^{-4}$, due to the
fact that four $e/4$ QPs are forced to tunnel simultaneously, in
agreement with earlier findings that electron tunneling is least
relevant in the MR state.\cite{fendley_edge_2007}
Figure~\ref{fig:fermi-e/4-e-cond} shows the differential  conductance,
$dI/dV$, in the cotunneling regime, for electron tunneling in the FL
regime and in the MR state via a QD, and for $e/4$ and $e/2$ QP
tunneling in the MR state via an AD at experimentally achievable
temperatures, (a) $T=\Delta/100 k_B\approx1$ mK and  (b) $T=\Delta/20
k_B\approx5$ mK, respectively.  In the FL regime, which corresponds to
noninteracting excitations of the FQH edge, previous results
\cite{recher_quantum_2000,golovach_transport_2004} are recovered.
Fig.~\ref{fig:peak-temperature-dependence} shows the  temperature
dependence  of the $e/4$ and $e/2$ QP cotunneling  conductance at the
transition from the elastic to the inelastic regime, $V=\Delta$. 
The inset gives both the $e/4$ QP conductance dip depth, 
$-dI/d\tilde V$ at the minimum with respect to $V$, and the 
temperature range for which the negative differential conductance of 
$e/4$ QPs is observed.  The MR state reveals its
special signature in the line shape of the cotunneling conductance.
For $e/4$ QP tunneling regions of negative differential conductance
appear.  On the other hand,  both $e/2$ and $e/4$ QPs show pronounced
conductance peaks at the opening of a new transport channel.  These
special peaks corroborate the findings of
Ref. \onlinecite{fendley_edge_2007} of $e/2$ and $e/4$ QP tunneling
being  relevant.  In the cotunneling regime, however, the
renormalization group flow is cut off by bias $V$ and temperature $T$
such that the perturbative result is accurate.  Our calculations show
that  the different charge carriers can be clearly distinguished by
standard transport measurements.

Our approach is also applicable to a setup of two separate FQH samples
with common Laughlin FQH edge states at filling factor $\nu$ weakly
coupled through a QD, similar to the setup in Fig. \ref{fig:setup}
(a).  The cotunneling current is then given by Eqs. \eqref{eq:I_el}
and \eqref{eq:I_inel} with $\kappa=1/\nu$.\cite{wen_topological_1995}
In this scenario, $1/\nu$ QPs of fractional charge $\nu$ combine to a
full electron charge when tunneling through the QD, such that not one
particle has to tunnel but $\frac1\nu-1$ additional particles.  Due to
the necessity of simultaneous tunneling the electron cotunneling
current is suppressed by $\Lambda^{2-2/\nu}$.  It is interesting to
note that electron tunneling via a QD in both the MR state and the
Laughlin state at filling factor $\nu=1/3$ show the same voltage
dependence.  However, the two states differ in the velocities of the
excitations due to the bosonic and fermionic nature of the lead
eigenmodes.

In conclusion, we have shown that electron cotunneling via a quantum
dot is strongly suppressed in the Moore-Read state compared to $e/4$
and $e/2$ quasiparticle cotunneling via an antidot.  The line shape of
the differential conductance reveals the special signature of the
Moore-Read state. Both the Moore-Read wave function can be verified
and the charge carrying excitation can be determined by measuring the
cotunneling current in the setups depicted in Fig. \ref{fig:setup}.

\section{Acknowledgments}
We thank M. P. A. Fisher, D. Stepanenko, and O. Zilberberg for useful
discussions.  We acknowledge support from the Swiss NF, and the NCCRs
Nanoscience and QSIT.  B.B. acknowledges support by the EU-FP7
project SE2ND (271554).

\bibliography{lib}

\end{document}